\documentclass[pra,aps,amsmath,twocolumn,showpacs,floatfix]{revtex4}
\usepackage{graphicx}
\usepackage{bm}
\usepackage{pdfpages}
\input{epsf}
\usepackage{color}

\begin{document}
\setlength{\arraycolsep}{1mm}
\newcommand{\ot}[0]{\otimes}
\newcommand{\bk}[2]{\langle#1|#2\rangle}
\newcommand{\kb}[2]{\ke{#1}\br{#2}}
\newcommand{\da}{^\dagger}
\newcommand{\ket}[1]{\left\vert#1\right\rangle}
\newcommand{\bra}[1]{\left\langle#1\right\vert}
\newcommand{\vet}[1]{\underline{#1}}
\newcommand{\sket}[1]{|#1\rangle}
\newcommand{\sbra}[1]{\langle#1|}
\newcommand{\av}[1]{\left\langle#1\right\rangle}
\newcommand{\proj}[1]{\ket{#1}\!\bra{#1}}
\newcommand{\ketbra}[2]{\ket{#1}\!\!\bra{#2}}
\newcommand{\braket}[2]{\langle{#1}|{#2}\rangle}
\newcommand{\ii}{\rm i}
\newcommand{\ee}{\rm e}
\newcommand{\dd}{\mathrm{d}}
\newcommand{\dr}{{\rm d }\vec r}
\newcommand{\re}{\mathrm{Re}}
\newcommand{\im}{\mathrm{Im}}
\newcommand{\id}{\boldsymbol{1}}
\newcommand{\idm}{\mathbb{I}}
\renewcommand{\max}[0]{\mathrm{max}}
\renewcommand{\min}[0]{\mathrm{min}}
\renewcommand{\emph}[1]{{\it #1}}
\renewcommand{\vec}[1]{\boldsymbol{#1}}
\newcommand{\ver}[1]{\boldsymbol{\hat #1}}
\newcommand{\beq}{\begin{equation}}
\newcommand{\eeq}{\end{equation}}
\newcommand{\xb}{\mathbf{x}}
\newcommand{\rb}{\mathbf{r}}
\newcommand{\anna}[1]{{\color{red} #1}}


\title {Vortex stability in Bose-Einstein Condensates in the presence of disorder} 
\author {Marisa Pons$^{1}$}
\author {Anna Sanpera$^{2,3}$}
\affiliation{$^1$Departmento de F\' isica Aplicada I, Universidad del Pa\' is Vasco, UPV-EHU, Bilbao, Spain.}
\affiliation{$^2$ICREA, Pg. Llu\'is Companys 23, 08010 Barcelona,
Spain.}
\affiliation{$^3$Departament de F\'isica, 
Universitat Aut\`onoma de Barcelona, E-08193 Bellaterra, Spain.}

\date{\today}

\begin{abstract}
 
We study the time evolution of a Bose-Einstein Condensate with a vortex on it, when it is released from a trap and expands freely in an spatially uncorrelated disordered media. As customary in such cases, we perform the evolution over different disorder realizations and average over the disorder to obtain the quantities of interest. The propagation of non-interacting quantum systems in disordered media is strongly linked to Anderson localization which can be understood as formation of islands of constant phase due to coherent backscattering. We find that the vortex superfluid localizes in such media and, moreover, the vortex is resilient to disorder effects. This is a single particle effect. In the presence of interactions, no matter how small they are, the vortex rapidly decays into phase discontinuities although localization is still present.
The study of dispersion of a bosonic condensate with vorticity in a disordered media bears similarities with the stability of topological excitations in 2D p-wave fermionic superfluids where the ground state is a Majorana mode that arises in the form of a vortex in the order parameter. 

\end{abstract}

\date{\today}

\noindent{\pacs {03.75.Hh, 03.75.Lm, 67.85.-d}}

\maketitle
\vskip0.5cm

\section{Introduction}
Disorder effects are ubiquitous in Nature and strongly determine subtle features such as insulating properties of materials, the striking behavior of high-Tc superconductors or the inhibition of transport and the presence of localization. Single particle quantum localization can be formally defined, for unbounded systems, as
the phenomenon in which quenched static disorder on a single non-interactive particle Hamiltonian causes all its eigenstates to localize in space. Localization in this context emerges as a consequence of the coherent back scattering that appears when a wave packet spreads in the disordered media \cite{Anderson58}. In such cases, there are multiple interference paths which can add constructively. Those constructive paths, when averaged over the disorder, lead to an exponentially localized wave packet. The spectrum of the Hamiltonian is no longer continuous, but rather becomes dense point-like; 
the eigenvalues are infinitely close to each other and the eigenfunctions are exponentially localized and characterized by a localization length $\xi$, see e.g., \cite{Lewenstein12} and references therein. Only in 1D has been rigorously demonstrated that no matter how small the disorder is, it always leads to an exponential localization of all the eigenfunctions \cite{Anderson58,Mott68,Borland63}. 

In 2D, it is expected that localization occurs also for arbitrarily small disorder, but its character interpolates smoothly between an algebraic profile for weak disorder, and an exponential one for strong disorder. In 3D, scaling theory predicts that there is a transition between diffusion and localization. The critical value at which this transition occurs (at $T=0$) is the so-called mobility edge $E_c$ \cite{Mott68} which indicates that low energy states, $E< E_c$, are exponentially localized while those with $E>E_c$ are extended. Thus, if all occupied states lie below the mobility edge, full localization appears. Recent experiments with ultracold atoms in speckle potentials where the interaction energy is tuned via Feshbach resonances, and the energy of the states via the depth of the optical potential, have provided a clear signature of the mobility edge \cite{Kondov11,Aspect12,McGehee13,Semeghini15}.  While the above description applies to non-interacting or single particle wave functions, it has recently been shown that also interacting systems might localize \cite{Basko06,Nandkishore15}. Many-body localization (MBL) refers to the lack of thermalization presented by some low dimensional strongly interacting systems at finite temperature. Such MBL states are associated to a dynamical quantum phase transition. Also, it has been argued that the critical disorder strength depends on the energy density, yielding to the so-called many-body mobility edge, but its existence and properties are still highly discussed (see e.g. \cite{DeRoeck16,Singh16}).

Localization and transport inhibition can hardly be regarded as the explicit features of Anderson quantum localization \cite{Anderson58}, since classical localization presents also such properties \cite{Piraud11}. Classical localization occurs either when a particle of energy $E$ is surrounded by a disordered landscape with hills of higher energy, or, for some models of disorder, by effect of percolation \cite{Aharony94}. In the former case, the classical localization length, defined as the averaged size of the classically allowed paths, increases with the energy of the particle $E$. In the latter case, the disordered is defined by a random variable in the degree of connectivity, setting an energy threshold, the so-called percolation threshold, above which particles with energy larger than the threshold are allowed to spread arbitrarily far. Quantum percolation adds interference effects to the percolation threshold. However, in quantum localization, a proper tailoring of the disorder correlation function can revert this feature and higher energy particles can present a lower localization length \cite{Piraud13a}.

To date, most studies of quantum localization in ultracold gases have exhaustively analyzed the onset of localization for the ground state of a trapped gas expanding freely in a disordered medium. Seminal experiments have demonstrated unambiguously quantum localization effects  by expanding a trapped BEC in a 1D wave guide with a very weak speckle potential superimposed on it, ensuring that interactions are highly inhibited \cite{Billy08}. There, it was shown that  in the presence of disorder, each matter wave localizes leading to a density profile with exponentially decaying tails and a localization length corresponding to that of non-interacting particles with momentum $1/\ell_{heal}=1/\sqrt{8 \pi n a_0} $, being $\ell_{heal}$ the healing length of the condensate, $n$ its density and $a_0$ the s-wave scattering length \cite {Dalfovo96}. The localization length as a function of the strength of the disorder in this experiment is consistent with the existence of a cross-over from exponential to algebraic localization, and therefore, with the theoretical prediction \cite{sanchezpalencia07} of the existence of an effective mobility edge at $k = 1/L_{dis}$, being $k$ the atomic wave vector and $L_{dis}$ the correlation length of the disorder. In \cite{Roati08}, a very different route was taken to achieve quantum localization. First, particle interactions were suppressed by means of  Feshbach resonances, and then the condensate was left to expand in a quasi periodic lattice reproducing the Aubry-Andr\'e model\cite{Aubry80} or the non-interacting Harper model \cite{Harper55} of quantum localization.  Also, the effects of disorder on collective excitations have widely been addressed \cite{Aspect09,Modugno10}.

Here, we are specifically interested in the behavior of a superfluid supporting topological charges, such as vortices, in the presence of quenched spatial disorder. With this aim in mind, we start by analyzing the onset of localization of excited states of 1D and 2D trapped Bose-Einstein Condensates expanding in the presence of a Gaussian uncorrelated random disorder. Current experiments using holographic masks allow implementing such type of disorder as well as the random impurity model or many other models in which the disorder is uncorrelated \cite{Bark09}.

Our paper is organized as follows: first we analyze the onset of localization when the initial state is an excited state of a trapped 1D condensate suddenly released on an uncorrelated disordered medium.  We calculate the localization length for a fixed ratio between the disorder strength and the initial state energy, observing that the localization length decreases as the energy of the initial state increases. In Sec. III, we move to the 2D case to analyze the interplay between vorticity, disorder and interactions. First, we examine the dispersion of a non-interacting state with a vortex in a disordered potential, showing that the vortex is resilient to disorder while the wave packet localizes, demonstrated by a quantized circulation along a closed path. As soon as interactions are included, no matter how small they are, the vortex signature disappears, even though the wave packet presents localization, indicating that interactions push the vortex to the borders of the condensate before localization takes place. Previous models analyzing dispersion of topological defects like bright solitons in spatially correlated disordered media have shown the presence of Anderson localization \cite{Krak09}. Our study of the vortex stability can also be applied in e.g. 2D p-wave superfluids that have a zero energy excitation (Majorana mode) in the form of a vortex \cite{Reed}.

\section{1D case}

Previous studies \cite{sanchezpalencia10} in 1D and 2D BEC have shown that the localization length,
$\xi$, is linked to the Fourier components of the disorder potential at typical wave vector $k_{E} =\sqrt{2mE}/\hbar$, where $m$ stands for the particle mass and $E$ for its typical energy.  Also, it has been shown that in 1D $1/\xi(E)\propto \tilde{C}(2k_{E})/E$ where $\tilde{C}$ is the correlation function of the disordered potential in Fourier space  $C(x)=\langle V_{dis}(x+x')V_{dis}(x')\rangle-\langle V_{dis}\rangle^{2}$ \cite{Piraud13a}. For uncorrelated disorder, $\tilde C$ is just a constant that has no dependence whatsoever on the momentum but depends on the strength of the disorder.

The dynamics of the condensate at $T=0$ is usually described by the Gross-Pitaevskii equation (GPE), see for example \cite{Pitaevskii03,Pethick}:
\begin{equation}
\label{GPE}
i\hbar\frac{\partial \varphi}{\partial t}= \left[-\frac{\hbar^2}{2m} \nabla^2 +V_{trap} + V_{dis,i} +g|\varphi|^2 \right]\varphi
\end{equation}
where $m$ is the atomic mass and $g$ is the effective interaction strength that depends on the scattering length, $a_0$, the number of atoms, $n_{atom}$, and on the transverse confinement frequency, $\omega_\perp$. In 1D $g=2 a_0 \hbar \omega_\perp n_{atom} $. 
The GPE describing an interacting condensate in a disordered media presents some limitations since quantum fluctuations due to multiple scattering are not taken into account. However, it has been well established that for very  weakly interacting systems such description is accurate. Indeed, in \cite{Stasinska12} we made a detailed analysis of the performance of the GPE compared to the more accurate Multi-Orbital Hartree-Fock method in an uncorrelated random potential, demonstrating that for weakly interacting systems, like the one we present here, both treatments lead to very similar results. On the contrary, in the case of many-body localization which may appear for some strongly correlated systems, the GPE cannot be used.

In what follows we consider the atoms to be $Rb^{87}$ and initially we assume non-interacting particles, i.e. $g=0$. At the end of this Section we comment on the results for $g\ne0$ obtained for this system. 
We assume the condensate to be trapped in a harmonic potential, $V_{trap}=\frac 12 m \omega^2 x^2$ 
where  $\omega$ is the trapping frequency. The disorder potential $V_{dis}$ is taken to be spatially uncorrelated with a Gaussian distribution:
\begin{equation}
V_{dis,i}(x,t)= \left\{ \begin{array}{rl}
   0, &\mbox{$t=0$} \cr
  Dn_{i} (x), & \mbox{$t > 0$}
       \end{array} \right.
       \label{vdis}
\end{equation}
where $n_{i}(x)$ is the Gaussian random distribution (zero average and unit variance) for the $i$-th realization of the disorder, and $D$ is the strength of the disorder. Our numerical simulations of the GPE are based on finite differences combined with the split-operator method and the results are obtained after averaging the quantities of interest over the $N$ different random realizations of the disorder potential $V_{dis,i}$.
\begin{figure}
\centering
\includegraphics[width=8.8cm,height=6cm]{{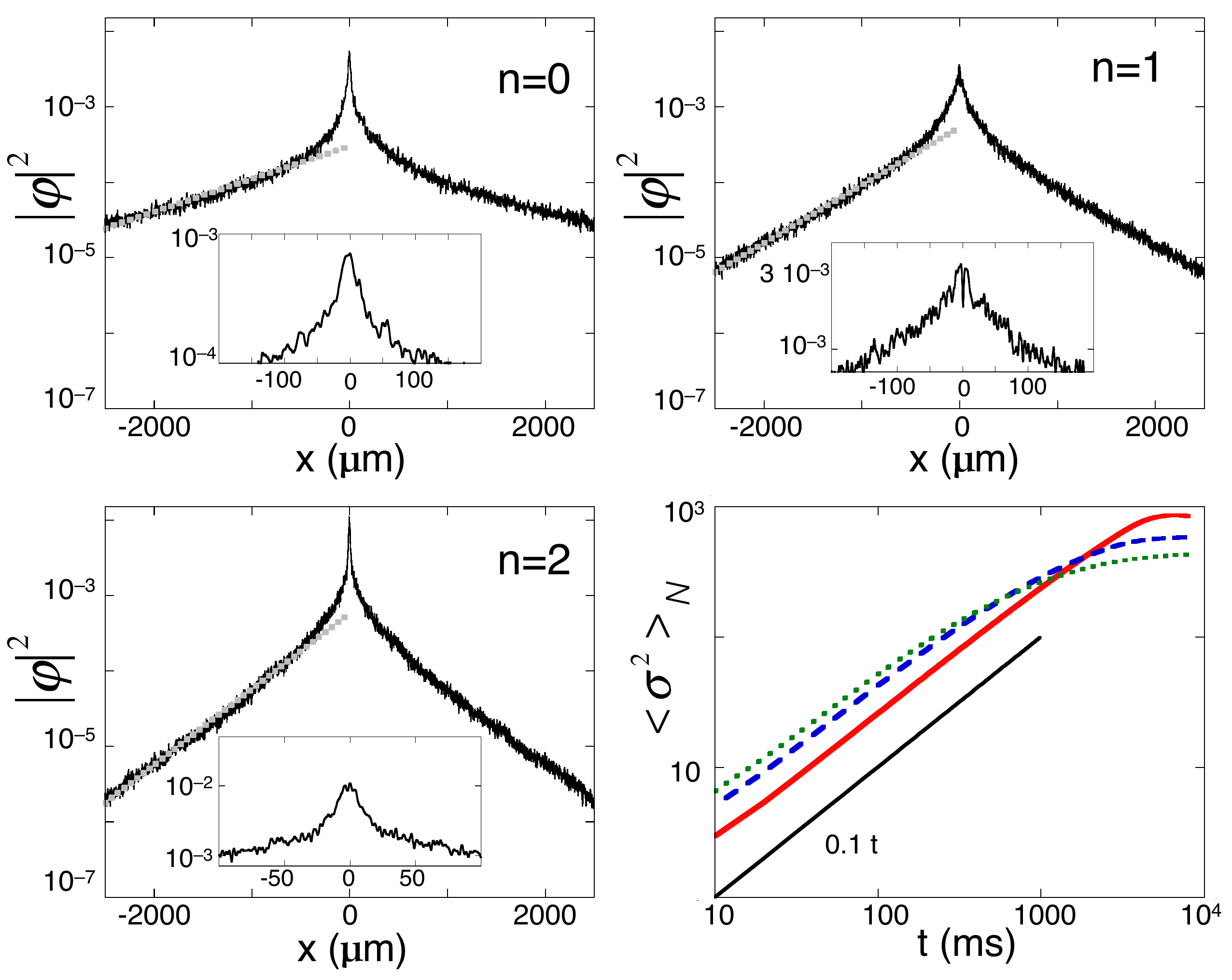}}
\caption{(Color online) Density of the wave funcion $|\varphi_{n}(x,\tau)|^2$ (solid line) after diffusing for a time $\tau=8 s$. The system is initially prepared in the state $\varphi_n(x)$ with $n=0$ (top-left), $n=1$ (top-right) and $n=2$ (bottom-left). The dashed lines are the best exponential fit functions $\exp[-x/L_n]$ (see text). In the insets the rescaled initial density distributions $|\varphi_n(x,\tau)|^2$ near $x=0$ are shown. Bottom-right panel: variance $\langle\sigma^2\rangle_N$, in $\mu m^2$, averaged over the disorder realizations as a function of time for the initial state $n=0$ (thick solid), $n=1$ (dashed), $n=2$ (dotted). The numerical data are compared to a simple linear function $0.1 t$ describing diffusion. }
\label{figure1}
\end{figure}


At $t=0$, the system is prepared in one of the eigenstates of the Hamiltonian (\ref{GPE}) (i.e., $V_{dis,i}=0$) with
\begin{eqnarray}
\varphi(x,0)&=&\varphi_{n}(x)\quad n=0,1,2,\dots\nonumber\\
\varphi_{n}(x)&\equiv& 
\frac{1}{\sqrt{2^n n!}}
\left(\frac{1}{\pi\lambda^2}\right)^{1/4} 
e^{-\frac{x^2}{2\lambda^2}}H_n\left(\frac{x}{\lambda}\right)
\label{eq:osc0}
\end{eqnarray}
where $H_n$ are the Hermite polynomials and $\lambda=\sqrt{\hbar/m\omega}$ is the oscillator length.

For the non-interacting case, we take the strength of the disorder $D$ to be proportional to the energy of the initial eigenstate of the harmonic potential, i.e. $d=D/E_{n}=0.52$. This permits a straight comparison between the localization length of the different initial states. To ensure that classical localization is avoided we further bound the disorder strength such that $V_{dis,i}=0$ either if $D n_{i} (x)>0.7 E_{n}$ or $V_{dis,i}<0$. The number of disordered potentials used for the averages is $N=100$, since the results converge already for this value of $N$.  At $t > 0$, the trapping potential is suddenly switched off while the disordered potential is turned on. To signal the onset of localization, we allow the wave packet to evolve in time until the dispersion of the initial wave packet 
averaged over the disorder realizations remains constant, i.e.
$\langle{\sigma}^2\rangle_{N}=\langle{\langle{x^2}\rangle-\langle{x}\rangle^{2}}\rangle_{N}$ 
does not depend on time. The numerical simulations show a ballistic expansion for very short times followed by a clear diffusive behavior $\langle{\sigma^2}\rangle_{N} \approx t$ finally leading to a steady situation for longer times $\tau$. When this stage is reached, we monitor the wave packet density distribution and observe that its shape remains constant for longer times. 


As shown in Fig. \ref{figure1}, the final density $|\varphi_{n}(x,\tau)|^2$, averaged over the disorder, shows exponential tails, while the central part of the localized wave function reproduces the nodes of the initial eigenstate $|\varphi_{n}(x)|^2$, as displayed in the corresponding insets. This is witnessed by the number of peaks around $x=0$ as compared to the initial wave function. An exponential ansatz of the form $|\varphi_n(x,\tau)|^2=A_n \exp[-x/\xi_n]$ gives the best fit results. For $n=0,1,2$ we obtain the following localization lengths $\xi_0=1300 \mu m; \xi_1 = 580 \mu m; \xi_2 = 440 \mu m$, indicating that $\xi_{n}$ decreases with increasing energy.


If we repeat the simulations for a much higher value of the disorder potential, so that $D\gg E_{n}$, several features remain, as for instance, the shape of the wave function near the origin and the exponential tails of the wave function.  Localization occurs now at a much shorter time scale but the localization length $\xi_{n}$ becomes independent of the initial eigenstate, evidencing that classical localization takes place. In such cases, an ansatz of the form $|\varphi_{n}(x)|^{2}=B_n x^{3/2}\exp[-x/S]$  gives a good fit of the overall wave function \cite{Palpacelli08}, where $S$ plays now the role of a localization length and its value is, to all effects, independent on the initial state $n$.  Finally, we have analyze interacting condensates, parametrized by a two-body interaction $g\neq 0$. Not surprisingly, the above effects are no longer visible.

\section{2D Case}

As it is well recognized \cite{Miniatura09,Pezze11,Piraud13b}, the dynamical behavior of even non-interacting particles in disorder potentials is a difficult problem in dimensions higher than one. 
A detailed analysis of classical localization, with diffusive, superdiffusive and subdiffusive behavior has been
recently presented for random correlated disorder in the regime where  
$\lambda_{dB}\ll \sigma_{r}\le l_{B}\ll L$ where  $\lambda_{dB}$ corresponds to the atomic de Broglie wavelength, 
$\sigma_{r}$ is the correlation length of the disorder, $l_{B}$ is the mean free path (Boltzmann) and $L$ the size of the system. Under the above circumstances, it has been shown \cite{Pezze11} that a variety of diffusion regimes appear depending on the disorder properties.
In the 2D case, the general Hamiltonian writes,
\begin{equation}
\label{GPE2}
i\hbar\frac{\partial \varphi}{\partial t}= \left[-\frac{\hbar^2}{2m} \nabla^2 +V_{trap} + V_{dis,i} 
+g_{2D}|\varphi|^2-L_{z}\Omega \right] \varphi.
\end{equation}

The trapping potential is $V_{trap}(\rb)=\frac 12 m \omega^2r^2$ with equal trapping frequencies $\omega=\omega_{x}=\omega_{y}$ and the condensate is rotating around the $z$ axis at an angular velocity $\Omega$.  The spatially uncorrelated disordered potential, $V_{dis,i}(\rb,t)=Dn_{i} (\rb)$, is switched on at $t > 0$ while the trapping potential and the rotation are switched off. 
When the angular frequency, $\Omega$, is large enough, the system minimizes the energy by means of vortices.
For the non-interacting case, we consider the following stationary states of the rotating Hamiltonian: $\varphi_{0,0}(\rb)=\exp{(-r^{2}/2\lambda^2)}/\sqrt{\pi \lambda^2}$ with $r^{2}=x^2+y^2$ corresponding to $n=l=0$ and the state 
$\varphi_{1,1}(\rb)=(x+i y) \exp{(-r^2/2 \lambda^{2})}/\sqrt{\pi\lambda^2}$ ($n=l=1$) with a vortex in it. When interactions are included, we obtain the stationary states numerically from Eq.(\ref{GPE2}) using imaginary time evolution with $V_{dis,i}=0$.

\begin{figure}
\includegraphics[width=16cm]{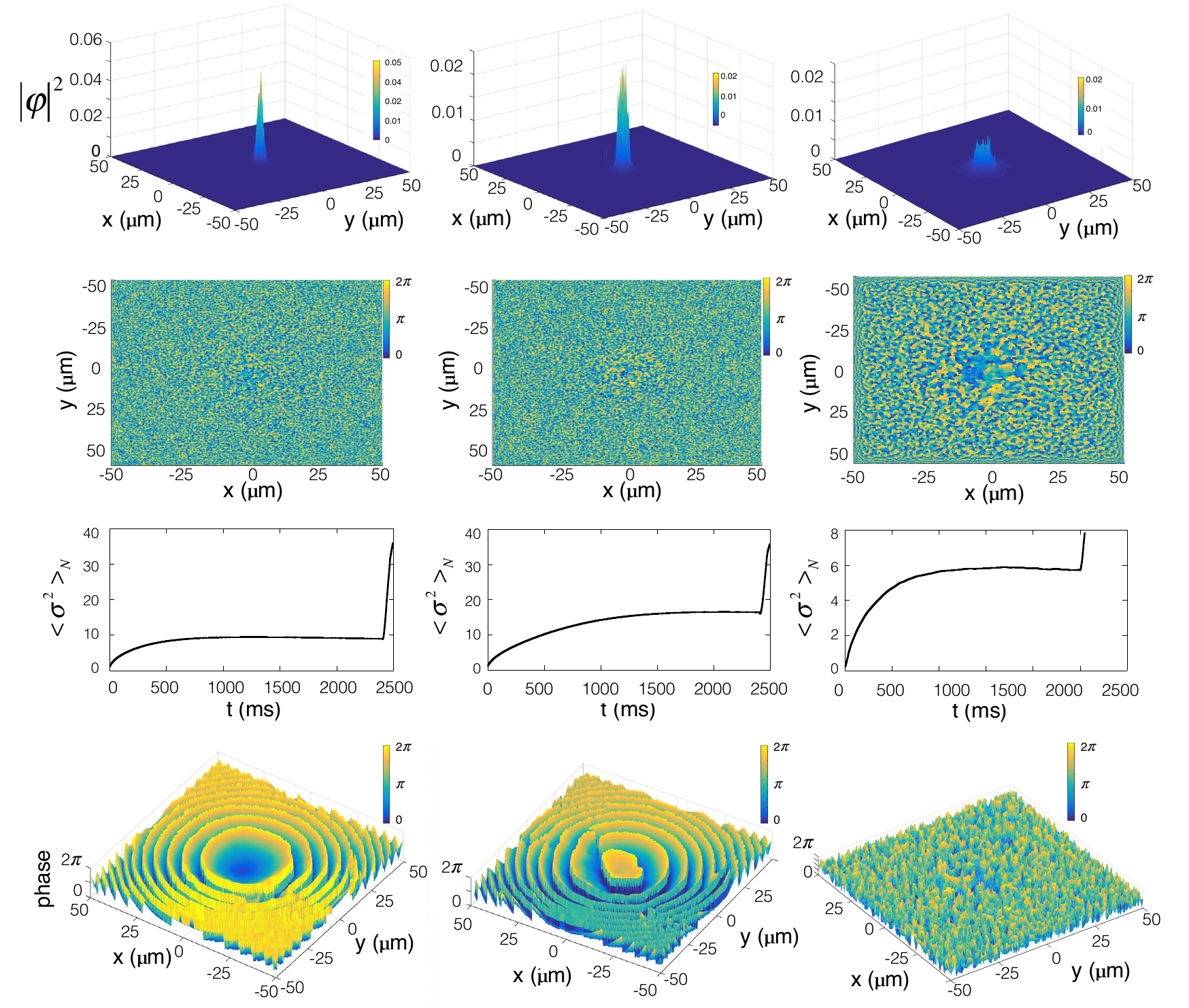}
\onecolumngrid 
\caption{(Color online) The left column corresponds to the results for the initial state $\varphi_{0,0} (x,y)$, the middle column to the non-interacting condensate prepared in the excited state $\varphi_{1,1} (x,y)$ while the right column shows the evolution of a weakly interacting condensate $g_{2D}=3.5 \times 10^{-24}$ corresponding to $10^3$ atoms with a vortex on it. All the figures display the averaged quantities over the disorder realizations. From top to bottom, (i) wave function density after expanding freely in the presence only of the disordered potential at $t=2.45s$, (ii) phase of the wave function, in radians, at $t=2.45s$, (iii) variance $\langle \sigma^2 \rangle_{N}$ of the position, in $\mu m^2$, as a function of time, (iv) phase of the wave function at $t=2.5s$ after switching off the disorder and let the condensate evolve freely for 5ms.} 
\label{figure2}
\end{figure}

The procedure we follow is as before: for the ideal gas ($g_{2D}=0$), at $t=0$ we take as initial wave function either $\varphi_{{0,0}}$ or $\varphi_{1,1}$ in the absence of disorder and relase the wave packet from the harmonic trap, switching off the rotation and quenching the disordered potential $V_{dis,i}$. The later is modeled again by an uncorrelated Gaussian distribution of strength $D$. In the 2D case, there is no need to limit the strength of the disorder to lower values than the energy of the eigenstate since classical localization is not guaranteed for uncorrelated disordered in such geometry. Therefore, we fix $D/E_{n}=6$ where $E_{n}$ is the energy of the initial state in the trap. Simulations in 2D for smaller disorder strength are accompanied by much longer times to achieve transport suppression, which, in turn demands prohibitively large grid sizes. We use $N=60$ different disorder realizations for each initial state and calculate the averaged values of the density wave function, its phase and the variance of the wave function. When the variance remains constant over time, so that the diffusive character of the initial expansion has ended, we compute the density of the BEC and its phase and average our results over all disorder realizations. For the interacting case, $g_{2D}=({8\pi \hbar^3 \omega_z/m})^{1/2} a_0 n_{atom}$ \cite{Dum}, and we fix the number of atoms to $n_{atom}=10^3$ so that we simulate a weakly interacting gas. In Fig. \ref{figure2}, we present our results for ideal gas in the ground state, $\varphi_{0,0}$ (left column), the ideal gas with a vortex state $\varphi_{1,1}$ and the interacting gas with a vortex on it.

\onecolumngrid

\twocolumngrid

%
%

As clearly depicted by the averaged variances, in the third row of Fig. \ref{figure2}, the wave packet diffusion reaches a steady state for sufficiently long times in all cases. At $t=2.45 s$  we perform the average over the disorder realizations of all quantities. The density of the wave function, shown in the upper panels of Fig. \ref{figure2}, presents localization in all cases, for the ideal gas in the ground and the excited state as well as for the very weakly interacting case (right column) while the averaged phase of the BEC at such (or previous) times displays a random pattern in all cases as shown in the second row of Fig. \ref{figure2}. However, if we switch off the disorder and let the BEC evolve freely, the localized wave packet starts to expand (as shown by the variances, see Fig. \ref{figure2}) but the phase of the ideal gas, when averaged over the disorder, recovers the phase of the initial state.  Our main results are shown in the lower panels of Fig. \ref{figure2}. After the disorder has been switched off, the phase of the condensate initially in $\varphi_{0,0}$,  reproduces the propagation of an initial Gaussian state that is let to evolve freely, displaying a typical self-similar solution. The phase of the wave functions expands from zero to $2\pi$ centered at $x=y=0$, The rings show the $2\pi$ modulus of the phase. However, the density of the wave packet after the disorder has been switched off is not anymore Gaussian. Propagation of the ideal gas in the excited state ($n=1,l=1$) shows a similar effect. The initial state with a vortex on it localizes and presents a completely random phase in the presence of disorder (middle column in Fig. \ref{figure2}). As soon as the disorder is switched off and the wave function is left to evolve freely for 5ms, the phase arranges and presents a plateau of constant phase followed by a clear discontinuity on the phase near the origin, with a helicoidal structure on it, reflecting the rotation of the condensate. The presence of the vortex after the disorder has been switched off can be proved by calculating the irrotational velocity. Expressing the wave functions of the initial excited state at the final time in terms of a fluid density ${\rho(\rb,t)}$ and the macroscopic phase $S (\rb,t)$ via $\varphi(\rb,t) =  \sqrt {\rho(\rb,t)} \exp[iS(\rb, t)]$, the presence of a vortex can be assured when the wave function remains single valued if the change in phase around any closed contour $C$ is \cite{Fetter,Parker}

\begin {equation}
\Gamma= \frac{\hbar}{m}\int_C{\nabla S \cdot d{\bf l}}=2\pi q,
\label{circul}
\end {equation}
where $q$ is an integer. Applying Eq. (\ref{circul}) for each disorder realization and averaging over all realizations, we obtain $q\simeq 1$, as it happens at $t=0$ for this state. This demonstrates that the vortex is resilient to the propagation in the uncorrelated disorder. This, somehow surprising result, can be understood invoquing the Kelvin theorem circulation, which ensures that under conservative potentials vorticity should be preserved \cite{Pitaevskii03}. In the presence of interactions, however, the above picture breaks down completely. There, one cannot recover the phase corresponding to the initial wave function with a vortex on it, equivalently, Eq. (\ref{circul}) does not hold for $t>0$ and the average phase of the interacting condensate presents a random pattern at any time, as shown in Fig. \ref{figure2}, right column. The effect of interactions on the expansion of the condensate in the disordered media is twofold, first it pushes the vortex to the boundaries of the condensate during the ballistic expansion where the density of the wave packet is negligible and suppresses the phase coherence.  After this initial stage of ballistic free expansion, interactions become irrelevant and the gas behaves similarly to an ideal gas displaying localization.

\section{Summary}

We have presented here a numerical study on the effects of a random uncorrelated disorder in non trivial initial states. We have analyzed the onset of localization, both classical and quantum for a 1D gas with and without interactions, for different initial states. We have observed that for such model, the averaged density keeps memory of the initial state in which it was prepared and shows the typical exponential tails. If the disorder is forced to be always smaller than the energy of the initial state to avoid classical confining, the exponential length of the localized state decreases with the energy of the initial state. For classical localization this is not the case. Introducing interactions between particles, these features disappear. In 2D, we have focused on the stability and localization of a superfluid vortex state in uncorrelated disordered media. In this case we observe that if interactions are absent the wave function localizes independently of the presence of the vortex. For the ideal case, the phase of the initial wave function is recovered after propagation, once the disordered potential is turned off. When interactions are included, even if they are very small, such effect is no longer visible, at least in our a model of Gaussian uncorrelated disorder, indicating that vorticity is resilient to disorder as long as interactions are absent.

\section {Acknowlegments}

 We are clearly indebted to Valentina Caprara and Gabrielle de Chiara with for useful discussions discussed. We acknowledge financial support from the Spanish MICINN (Grant  No. FIS2013-40627-P, FIS2015-67161-P), the Generalitat de Catalunya AGAUR, contract 2014SGR-966, the  Basque  Government  (Grant No. IT-472-10), ERC starting grant GEDENTQOPT.


\begin{thebibliography}{10}
\bibitem{Lewenstein12} M. Lewenstein, A. Sanpera A, and V. Ahufinger, 
{\it Ultracold atoms in Optical Lattices: mimicking condensed matter and beyond} (Oxford University Press, Oxford) (2012). 
\bibitem{Anderson58} P.W. Anderson Phys. Rev. {\bf109}, 1492 (1958). 
\bibitem{Mott68} N.F. Mott, Rev. Mod. Phys. {\bf40}, 677 (1968). 
\bibitem{Borland63} R.E. Borland Proc. R. Soc. Lond. A {\bf 247}, 529 (1963). 
\bibitem{Kondov11} S.S. Kondov, W.R. McGehee, J.J. Zirbel, B.DeMarco, Science {\bf334} (6052), 66-68 (2011).
\bibitem{McGehee13} W.R. McGehee, S.S. Kondov, W.Xu, J.J. Zirbel, and B. DeMarco, Phys. Rev. Lett. {\bf111}, 145303 (2013).
\bibitem{Aspect12} F. Jendrzejewski {\it et al.}, Nature Physics {\bf8}, 398–403 (2012).
\bibitem{Semeghini15} G. Semeghini {\it et al.}, Nature Physics
{\bf 11}, 554–559 (2015).
\bibitem{Basko06} D.M. Basko, I.L  Aleiner and B. Altshuler Ann. Phys., NY {\bf 321} 1126–205 (2006).
\bibitem{Nandkishore15} R. Nandkishore and D. A. Huse, Annu. Rev.  Condens Matter Physics, {\bf 6},031214 (2105).
\bibitem{DeRoeck16} W. De Roeck, F. Huveneers, M. Müller, and M. Schiulaz
Phys. Rev. B {\bf 93}, 014203 (2016).
\bibitem{Singh16} R. Singh, J. H Bardarson and F. Pollmann, New J. Phys. {\bf 18}, 023046 (2016).
\bibitem{Piraud11} M. Piraud, P. Lugan, P. Bouyer, A. Aspect, and L. Sanchez-Palencia, Phys. Rev. A {\bf83}, 031603 (2011).
\bibitem{Aharony94} A. Aharony and D. Stauffer, {\it Introduction to percolation Theory} (Taylor and Francis, London, 1994).
\bibitem{Piraud13a} M. Piraud and L. Sanchez-Palencia,  Eur. Phys. J. Spec. Top. {\bf 217}, 217, (2013).
\bibitem{Billy08}  J. Billy {\it et al.}, Nature {\bf453}, 89 (2008). 
\bibitem{Dalfovo96} F. Dalfovo, L. Pitaevskii, and S. Stringari, Journal of Research of the National Institute of Standards and Technology, {\bf101}, 537 (1996)
\bibitem{sanchezpalencia07} L. Sanchez-Palencia {\it et al.},  Phys. Rev. Lett. {\bf98}, 210401 (2007). 
\bibitem{Roati08}  G. Roati {\it et al.} Nature {\bf453}, 895 (2008). 
\bibitem{Aubry80} G. Andr\'e and S. Aubry, Ann. Isr. Phys. Soc. {\bf3}, 33 (1990). 
\bibitem{Harper55} P.G. Harper, Proc. Phys. Soc. Lond. A {\bf 68}, 874 (1955). 
\bibitem{Aspect09} A. Aspect and M. Inguscio Phys. Today {\bf62}, 30 (2009).
\bibitem{Modugno10} G. Modugno Rep. Prog. Phys. {\bf73}, 102401 (2010). 
\bibitem{Bark09}  W. S. Bakr, J. I. Gillen, A. Peng, S. F\"olling  and M. Greiner, Nature  {\bf 462}, 74, 2009.
\bibitem{Krak09} K. Sacha, C. A. M\"uller, D. Delande, and  J. Zakrzewski, Phys. Rev. Lett. {\bf103}, 210402 (2009).
\bibitem{Reed} N.Read and D.Green, Phys. Rev. B {\bf 61}, 10267 (2000).
\bibitem{sanchezpalencia10} L. Sanchez-Palencia and M. Lewenstein, Nature Phys. {\bf6}, 87 (2010).
\bibitem{Pitaevskii03}
L. Pitaevskii and S. Stringari, {\it Bose-Einstein condensation} (Oxford University Press, Oxford) (2003).
\bibitem{Pethick} 
C.J. Pethick and H. Smith,  {\it Bose-Einstein Condensation in Dilute Gases}
(Cambridge University Press, New York) (2008).
\bibitem{Stasinska12} J. Stasi\' nska, P. Massignan, M. Bishop, J. Wehr, A. Sanpera, and M. Lewenstein New J. Phys. {\bf14}, 043043 (2012).
\bibitem{Palpacelli08} S. Palpacelli and S. Succi, Phys. Rev. E {\bf77}, 066708 (2008).
\bibitem{Miniatura09} C. Miniatura, R. C. Kuhn, D. Delande, and C. A. M\"uller, Eur. Phys. J. B {\bf68}, 353 (2009).
\bibitem{Pezze11}  L. Pezz\'e {\it et al.}, New J. Phys. {\bf13}, 095015 (2001).
\bibitem{Piraud13b} M. Piraud, L. Pezz\'e, and L Sanchez-Palencia, New J. Phys {\bf15}, 075007 (2013).
\bibitem{Dum} Y. Castin and R. Dum, Eur. Phys. J. D {\bf7}, 399 (1999)
\bibitem{Fetter} A. Fetter, Rev. Mod. Physics, {\bf81}, 647 (2009)
\bibitem{Parker} N. G. Parker, B. Jackson, A. M. Martin, and C. S. Adams, in {\it Emergent Nonlinear Phenomena in Bose-Einstein Condensates}, 173-189 (2007).




\end{thebibliography}
\end{document}